\documentclass{ws-ijmpcs}

\pdfoutput=1

\def\trimmarks{}

\usepackage{amsmath}

\begin{document}

\markboth{M. Diehl}{Multiple hard scattering}

%
\catchline{}{}{}{}{}
%

\title{\hfill \textnormal{DESY 14-205} \\[3em]
  Multiple hard scattering and parton correlations in the proton}

\author{Markus Diehl}

\address{Deutsches Elektronen-Synchroton DESY, Notkestrasse 85,\\
  22607 Hamburg, Germany\\
  markus.diehl@desy.de}

\maketitle


\begin{abstract}
  This proceedings contribution gives a brief introduction to the
  theoretical description of double parton scattering and discusses
  several open problems.
  \keywords{proton-proton collisions, double parton scattering; scale
    evolution.}
\end{abstract}



\newcommand{\tvec}[1]{\boldsymbol{#1}}

\newcommand{\ms}{\mskip 1.5mu}

\newcommand{\half}{\tfrac{1}{2}}


\section{Introduction}	

\begin{figure}
\centerline{
\includegraphics[width=0.45\textwidth,%
  viewport=0 -17 445 130]{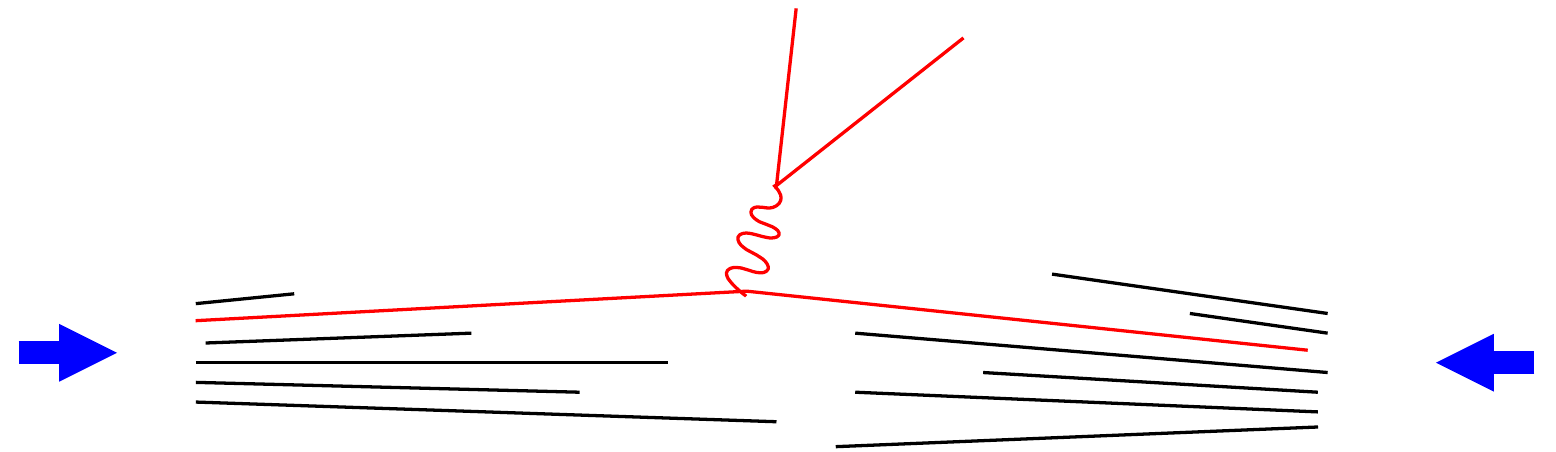}
\hspace{1em}
\includegraphics[width=0.45\textwidth]{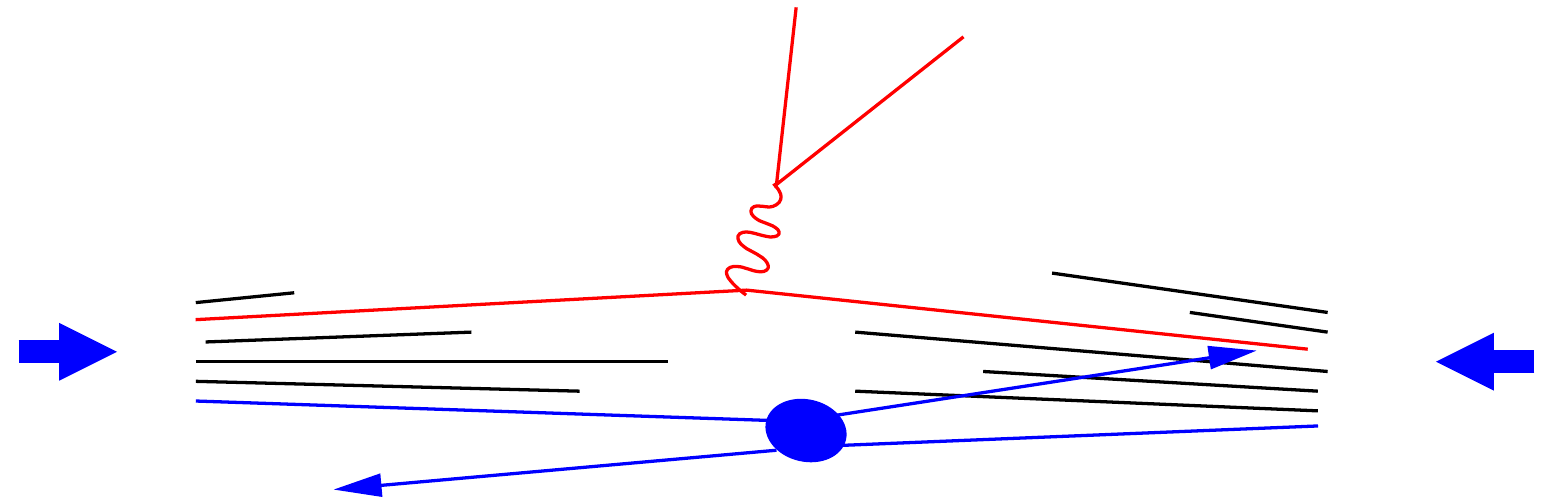}
}
\vspace*{8pt}
\caption{\label{f:dy-example} Sketch for the production and decay of a
  heavy gauge boson ($Z, W$ or off-shell $\gamma^*$) in a $pp$ collision.
  The left panel shows the picture suggested by the standard factorization
  formula for this process, whereas the right panel includes scattering
  between ``spectator partons'' that do not participate in the gauge boson
  production subprocess.}
\end{figure}

The standard description of hard processes in proton-proton (and other
hadron-hadron) collisions uses the concept of factorization.  In a
standard factorization formula, the proton-proton cross section is given
by the convolution of a parton density for each proton with a hard cross
section for the interaction of two partons, summed over the relevant
combinations of parton species.  The physical picture suggested by this
formula, represented in the left panel of Fig.~\ref{f:dy-example}, is
deceptively simple since it suggests that the only interaction taking
place in the collision is between the two partons initiating the hard
subprocess.  This is certainly not the case: the ``spectator partons'' in
each proton are colored and will interact with their counterparts in the
other proton, as sketched in the right panel of Fig.~\ref{f:dy-example}.

To solve this apparent contradiction, we first note that the factorization
formula describes semi-inclusive processes of the type $pp\to Y + X$.
Here $Y$ is the set of particles produced by the hard scattering (a lepton
pair in Fig.~\ref{f:dy-example}), whereas $X$ denotes all other
final-state particles.  One can specify all details of $Y$ (and compute
them from the hard scattering) but must sum over all possible hadronic
states in $X$.  The ``spectator'' interactions just mentioned affect
particles in $X$, and it is a non-trivial statement of the factorization
formula that---thanks to unitarity---one can ignore these interactions
when computing the inclusive cross section \mbox{$\sum_X \sigma(pp\to Y + X)$}.

At the high collision energies of the LHC and the Tevatron, ``spectator''
interactions can themselves be hard and produce particles with large mass
or large transverse momentum.  For many purposes it is then not sufficient
to known only $\sum_X \sigma(pp\to Y + X)$, and one is interested in the
details of $X$.  Recall that many search channels for new particles have
high multiplicity due to long decay chains; in this case some of the
relevant particles may belong to $Y$ but others to $X$.  A sufficiently
quantitative understanding of multiple hard interactions in $pp$
collisions is therefore important.

Even to understand double parton scattering, with two hard scatters as in
the right panel of Fig.~\ref{f:dy-example}, remains a challenge for theory
and phenomenology.  Building on pioneering work done in the
1980s,\cite{Paver:1982yp,Mekhfi:1983az} we have in
Refs.~[\refcite{Diehl:2011tt,Diehl:2011yj}] shown that several aspects of
double parton scattering allow for a systematic treatment in QCD, but that
a number of open issues remain to be understood.  This proceedings
contribution presents a selection of theory results and of open questions.


\section{Double scattering cross section}

\begin{figure}
\centerline{\includegraphics[width=0.5\textwidth,%
  viewport=0 0 440 250,clip=true]{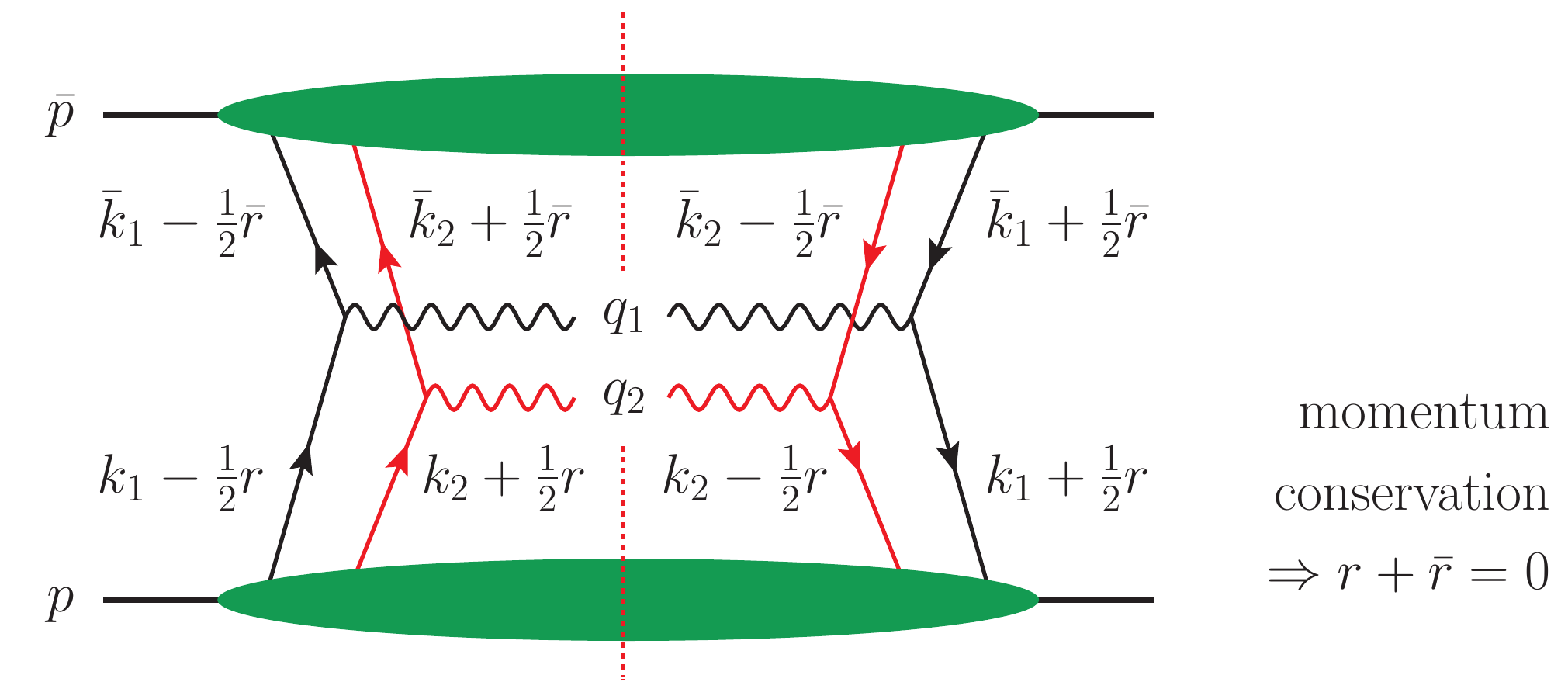}}
\vspace*{8pt}
\caption{\label{f:double-mom} Graph for the production of two gauge bosons
  (with momenta $q_1$ and $q_2$) by two independent hard-scattering
  processes.  Double parton distributions are represented by the blobs.
  The graph is for the inclusive cross section, with the final-state cut
  indicated by a dotted vertical line.  Using momentum conservation one
  easily finds $r + \bar{r} = 0$ for the momentum mismatch between partons
  on the left and the right of the cut.}
\end{figure}

Let us assume that proton-proton collisions in which two pairs of partons
initiate two independent hard-scattering processes can be described by a
factorization formula akin to the familiar one for single hard scattering.
The bound-state structure of each proton is then described by double
parton distributions, i.e.\ by the joint distributions of the two
scattering partons inside a proton.  A corresponding graph is shown in
Fig.~\ref{f:double-mom}.

A simple kinematic analysis shows that the plus- and minus-momentum
components of each parton in the graph are fixed by the observable gauge
boson momenta, because the partons in one proton carry only large
plus-momentum and those in the other proton only large minus-momentum.  By
contrast, all partons generically carry transverse momenta of similar
size.  As a consequence, the partons that produce a given boson can have
different transverse momenta in the scattering amplitude and in its
conjugate, and the cross section formula involves a transverse-momentum
convolution even if the transverse momenta of the gauge bosons are
integrated over.  In that case, the factorization formula can be written
as
\begin{align}
  \label{eq:coll-dpd-X}
\frac{d\sigma_{\text{double}}}{dx_1\ms d\bar{x}_1\,
  dx_2\ms d\bar{x}_2}
&= \frac{1}{C}\, \hat{\sigma}_1\ms \hat{\sigma}_2
\int \frac{d^2\tvec{r}}{(2\pi)^2}\;
F(x_1, x_2, \tvec{r}) \, F(\bar{x}_1, \bar{x}_2, -\tvec{r}) \,,
\end{align}
where $C$ is a combinatorial factor and $\hat\sigma_i$ is a
hard-subprocess cross section.  $F(x_1,x_2, \tvec{r})$ and
$F(\bar{x}_1,\bar{x}_2, -\tvec{r})$ are double parton distributions, whose
$x$ arguments are fixed as $x_i = (\bar{p}\ms q_i)/(\bar{p}\ms p)$ and
$\bar{x}_i = (p\ms q_i)/(p\ms\bar{p})$.  The formula is schematic in the
sense that the quantum numbers of the partons have been omitted and that a
sum over all relevant combinations should be taken.  More detail can be
found in \mbox{Refs.~[\refcite{Diehl:2011tt,Diehl:2011yj}]}.

Let us take the two-dimensional Fourier transform of $F(x_1,x_2,
\tvec{r})$ w.r.t.\ $\tvec{r}$.  The convolution integral in the cross
section then becomes
\begin{align}
\int \frac{d^2\tvec{r}}{(2\pi)^2}\;
F(x_1, x_2, \tvec{r}) \, F(\bar{x}_1, \bar{x}_2, -\tvec{r}) &=
\int d^2\tvec{y}\;
F(x_1, x_2, \tvec{y}) \, F(\bar{x}_1, \bar{x}_2, \tvec{y}) \,.
\end{align}
The Fourier conjugate variable $\tvec{y}$ can be interpreted as the
transverse distance between the two partons with longitudinal momentum
fractions $x_1$ and $x_2$, and hence as the transverse distance between
the two hard-scattering processes.

The result \eqref{eq:coll-dpd-X} is a collinear factorization formula at
tree level.  It can readily be generalized to include higher orders in the
subprocess cross sections $\hat\sigma_{i}$, which are identical to those
calculated for single hard scattering.  As in that case, the momentum
arguments of the parton distributions are then no longer fixed in the
cross section but appear in convolution integrals.

A different generalization of Eq.~\eqref{eq:coll-dpd-X} is for measured
transverse momenta $\tvec{q}_i$ of the gauge bosons.  If these are much
smaller than the hard scale $Q$ of the scattering (i.e.\ if $\tvec{q}_i^2
\ll q_i^2$) one has a TMD-type factorization formula, with double parton
distributions $F(x_1,x_2,\tvec{k}_1,\tvec{k}_2, \tvec{r})$ and
$F(\bar{x}_1,\bar{x}_2, \bar{\tvec{k}}_1,\bar{\tvec{k}}_2, -\tvec{r})$
(see Fig.~\ref{f:double-mom} for the momentum labeling).  The partial
Fourier transform $F(x_1,x_2,\tvec{k}_1,\tvec{k}_2, \tvec{y})$ of
$F(x_1,x_2,\tvec{k}_1,\tvec{k}_2, \tvec{r})$ has the structure of a Wigner
distribution in its transverse arguments: $\tvec{k}_1$ and $\tvec{k}_2$
are the transverse momenta of the two partons and $\tvec{y}$ is their
transverse distance if one ``averages'' over the two sides of the
final-state cut, i.e.\ over the scattering amplitude and its conjugate.

The factorization formula \eqref{eq:coll-dpd-X} and its generalizations
just discussed still have the status of conjectures: in
Ref.~[\refcite{Diehl:2011yj}] we could give several elements of a
factorization proof, but several issues remain to be clarified and worked
out.  Perhaps the most serious one is the question whether soft gluon
exchange between the spectator partons in the so-called Glauber region
(where gluon momenta satisfy $l^+l^- \ll \tvec{l}^2$) breaks factorization
or not.  The unitarity argument used for single gauge boson production
cannot be readily generalized to double hard scattering, and it remains to
be seen whether Glauber gluons invalidate the factorization
formula~\eqref{eq:coll-dpd-X}.


\section{Single vs.\ double hard scattering}

\begin{figure}[th]
\centerline{\includegraphics[width=0.85\textwidth,%
  viewport=160 325 460 575]{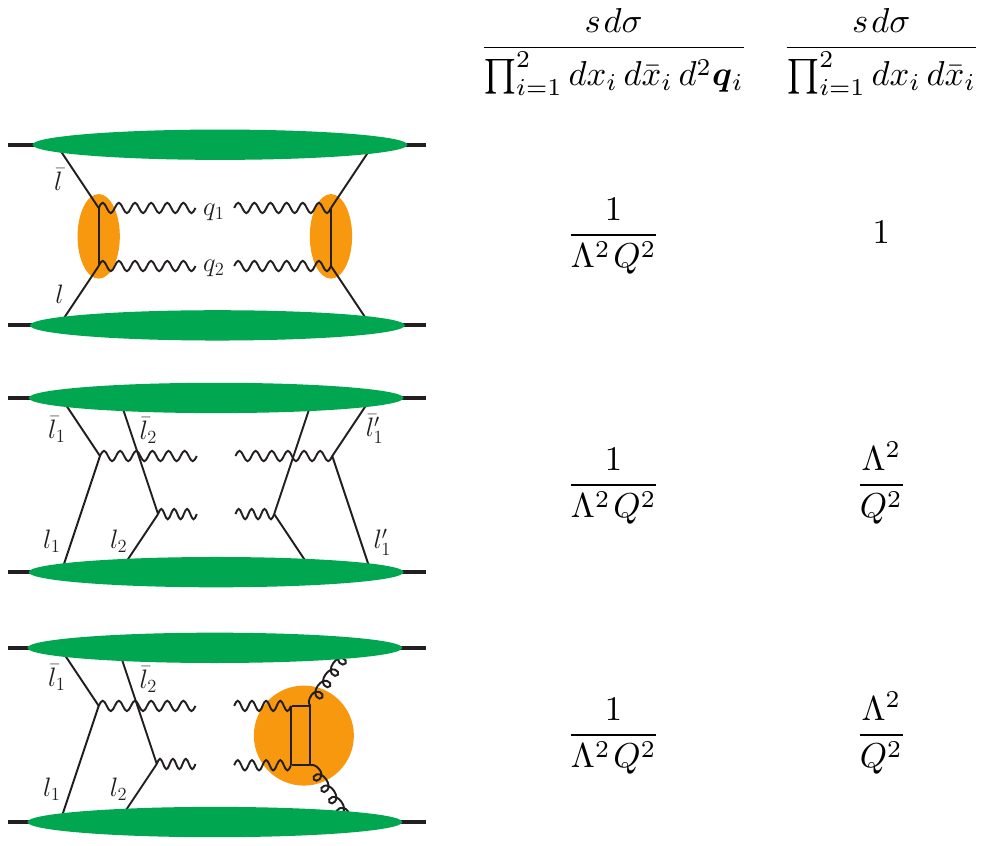}}
\vspace*{8pt}
\caption{\label{f:power-fig} Example graphs and power behavior for the
  production of two gauge bosons by single hard scattering (top), double
  hard scattering (middle) and their interference (bottom).  The shaded
  blobs indicate hard-scattering subprocesses.  $\Lambda$ denotes a
  generic soft QCD scale, and $Q$ the hard scale of the scattering.  The
  same power behavior is found for other final states.}
\end{figure}

A question of obvious concern is how important double hard scattering is
compared with the conventional single hard scattering mechanism.  The
answer depends crucially on the observed kinematic variables in the final
state.  As indicated in Fig.~\ref{f:power-fig}, the double scattering
mechanism is power suppressed by $\Lambda^2/Q^2$ w.r.t.\ single hard
scattering in the collinear factorization formula, but it is not
suppressed if the transverse boson momenta $\tvec{q}_1$ and $\tvec{q}_2$
are both measured.  For small transverse momenta, $|\tvec{q}_i| \ll Q$,
both mechanisms are thus generically of the same size, whereas in the
$\tvec{q}_i$ integrated cross section single hard scattering wins because
it populates a larger phase space with $|\tvec{q}_i| \sim Q$, which is
unaccessible to the double scattering mechanism.

At the same order in $\Lambda/Q$ as double hard scattering, there is also
the interference between single and double scattering, shown in the bottom
graph of Fig.~\ref{f:power-fig}.  In this case, the hadronic quantities
entering the collinear factorization formula are parton distributions of
twist-three, where all parton fields are located at the same transverse
position.  This is in contrast to double scattering, where the partons
associated with one or the other hard-scattering subprocess have a
relative transverse distance $\tvec{y}$ that is integrated over at the
level of the cross section and not of the individual parton distributions.
To our knowledge, such interference contributions have been ignored in
phenomenological considerations so far, and nothing is known about their
size in specific processes.

Apart from the hard scale $Q$, there is another parameter that controls
the relative size of contributions in a generic manner, namely the overall
$pp$ collision energy, which at given $Q$ translates into the typical size
$x$ of the momentum fractions of the participating partons.  Assuming the
absence of correlations between small-$x$ partons, one finds that the
cross sections for single and double parton scattering respectively scale
like\,\cite{Diehl:2011yj}
\begin{align}
\frac{d\sigma_{\text{single}}}{dx_1\ms d\bar{x}_1\,
  dx_2\ms d\bar{x}_2} &\sim \frac{1}{Q^2}\,
  x^{- 4 - 2\lambda} \,,
&
\frac{d\sigma_{\text{double}}}{dx_1\ms d\bar{x}_1\,
  dx_2\ms d\bar{x}_2} &\sim \frac{\Lambda^2}{Q^4}\,
  x^{- 4 - 4\lambda}
\end{align}
if the single-parton densities have a small-$x$ behavior $x f(x) \sim
x^{-\lambda}$.  Unless correlation effects overturn this trend, double
parton scattering thus becomes enhanced at low $x$.  A corresponding
enhancement is also found if one considers gluon initiated hard-scattering
processes and computes the energy dependence within the BFKL
framework.\cite{Braun:2000ua} Unfortunately, not enough is known about the
small-$x$ behavior of the three-parton correlation functions needed for
the interference term in Fig.~\ref{f:power-fig}, so that it remains
unclear whether it benefits from a similar low-$x$ enhancement.


\section{Parton correlations}

The double parton distributions $F(x_1,x_2, \tvec{y})$ (and even more so
their TMD counterparts) are barely known at present.  As a zeroth-order
approximation, one may assume the absence of correlations between the two
partons and then obtains
\begin{align}
  \label{eq:dpd-fact-y}
F(x_1,x_2, \tvec{y}) \approx
\int d^2\tvec{b}\;  f(x_1,\tvec{b}+\tvec{y})\, f(x_2,\tvec{b}) \,,
\end{align}
where $f(x, \tvec{b})$ is the impact parameter distribution of a single
parton,\cite{Burkardt:2000za} i.e.\ the probability density for finding a
parton with momentum fraction $x$ at a transverse distance $\tvec{b}$ from
the proton center.  In transverse-momentum space, this relation is even
simpler and reads\,\cite{Blok:2010ge}
\begin{align}
  \label{eq:dpd-fact-r}
F(x_1,x_2, \tvec{r}) \approx  f(x_1,\tvec{r})\, f(x_2,-\tvec{r}) \,,
\end{align}
where $f(x, \tvec{r})$ is the Fourier transform of $f(x, \tvec{b})$
w.r.t.\ $\tvec{b}$.

Several arguments speak against the independence of the two partons
expressed in \eqref{eq:dpd-fact-y} and \eqref{eq:dpd-fact-r} if the
momentum fractions $x_1$ and $x_2$ are not very small.  We will not go
into details here but instead refer to Ref.~[\refcite{Diehl:2013mma}].
Such correlations may concern only the overall size of $F(x_1,x_2,
\tvec{y})$, or depend only on the momentum variables $x_1$ and $x_2$, or
correlate $x_1$ and $x_2$ with $\tvec{y}$.

Moreover, the spin or the color of the two partons can be correlated.
Such correlations can be quantified by polarization or color dependent
distributions, which must be included in the factorization formula
\eqref{eq:coll-dpd-X}.  For these distributions, a factorized form as in
\eqref{eq:dpd-fact-y} and \eqref{eq:dpd-fact-r} is hardly plausible even
as a starting point.

Correlations between $x_1, x_2$ and $\tvec{y}$ can have important
quantitative consequences on double scattering cross
sections,\cite{Frankfurt:2003td} and the same is true for parton spin
correlations.\cite{Kasemets:2012pr} Color correlations are suppressed by
Sudakov factors in the collinear factorization
formula,\cite{Mekhfi:1988kj,Diehl:2011yj,Manohar:2012jr} but according to
the estimate in Ref.~[\refcite{Manohar:2012jr}] this suppression is not
very strong for moderately high scales.


\section{Short-distance behavior, evolution and an unsolved problem}

\begin{figure}[b]
\centerline{
\includegraphics[width=0.37\textwidth,
  viewport=0 -2 375 145]{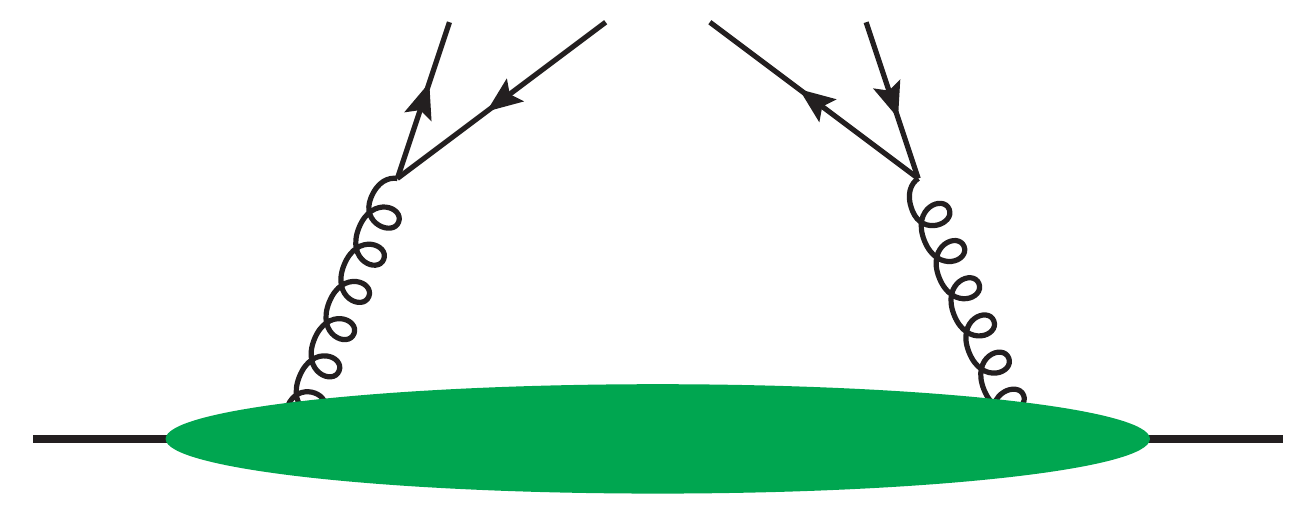}
\hspace{3em}
\includegraphics[width=0.4\textwidth]{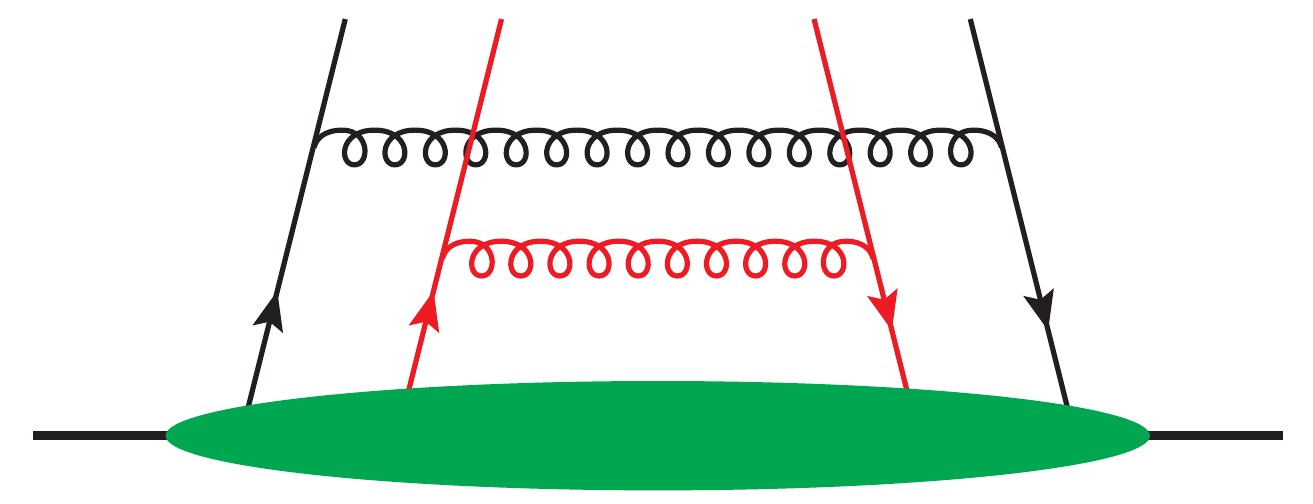}
}
\vspace*{8pt}
\caption{\label{f:evol} Left: graph for the splitting of one parton into
  two.  Right: ladder graph for the independent scale evolution of the two
  partons.}
\end{figure}

Using power counting arguments, one can show that for $|\tvec{y}| \ll
1/\Lambda$ the distributions $F(x_1,x_2, \tvec{y})$ can be computed in
terms of single parton densities and a hard-scattering kernel $K$ that
describes the splitting of one parton into two.\cite{Diehl:2011yj} An
example is shown in the left panel of Fig.~\ref{f:evol}.  At lowest order
in $\alpha_s$ one then has\,\cite{Diehl:2011yj}
\begin{align}
  \label{eq:dpd-small-y}
F(x_1,x_2, \tvec{y}) = \frac{1}{\pi}\,
  \frac{1}{\tvec{y}^2}\, \frac{f(x_1+x_2)}{x_1+x_2}\,
  K\Bigl( \frac{x_1}{x_1+x_2} \Bigr) \,,
\end{align}
whereas as higher orders one obtains a convolution over the momentum
fraction of the single-parton density.  In this perturbative regime one
finds strong spin correlations (the corresponding splitting kernels are
listed in Ref.~[\refcite{Diehl:2014vaa}]).  For the splitting $g\to
q\bar{q}$ shown in Fig.~\ref{f:evol}, the helicities of the quark and
antiquark are 100\% anti-aligned due to chirality conservation in the
massless-quark limit.

Let us now discuss the scale dependence of double parton distributions.
For the transverse-momentum dependent functions
$F(x_1,x_2,\tvec{k}_1,\tvec{k}_2, \tvec{y})$ one can derive Collins-Soper
evolution equations in close analogy to the case of TMDs, as shown in
Ref.~[\refcite{Diehl:2011yj}].  Their solution naturally provides the
Sudakov factors that resum large logarithms of $\tvec{q}_i^2 /Q^2$ in the
TMD factorization formula.  The evolution of collinear double parton
distributions with color correlations has been discussed in
Ref.~[\refcite{Manohar:2012jr}] and is again closely related with the
Sudakov logarithms we already mentioned in the previous section.

Collinear double parton distributions without color correlations follow
DGLAP-type evolution equations.  Two different versions of these have been
discussed in the literature.  It is natural to define the $\tvec{y}$ space
distributions for unpolarized partons as
\begin{align}
& F(x_1,x_2, \tvec{y}; \mu)
\nonumber \\
& \quad =  2p^+ \!\!
  \int  dy^-\,\frac{dz_1^- dz_2^-}{(2\pi)^2}\;
  e^{i p^+ (x_1^{} z_1^- + x_2^{} z_2^-)}\,
  \bigl\langle p \big| \ms
    \mathcal{O}(y,z_1;\mu)\, \mathcal{O}(0,z_2;\mu)
  \ms \big| p \bigr\rangle \,,
\end{align}
where
\begin{align}
\mathcal{O}(y,z;\mu) &= \tfrac{1}{2}\ms \bigl[
   \bar{q}\bigl(-\half z+y\bigr) \ms \gamma^+ q\bigl(\half z+y\bigr)
\bigr]_{\mu} \,,
& z^+ = y^+ = 0\,, \tvec{z}=\tvec{0}
\end{align}
are the familiar twist-two light-cone operators, renormalized at scale
$\mu$ in the same way as for single parton distributions.  We note in
passing that $F(x_1,x_2, \tvec{y}; \mu)$ involves the product of two
twist-two operators at a spacelike distance $y$ from each other, which is
to be distinguished from a twist-four operator.  The scale dependence then
follows a homogeneous DGLAP equation
\begin{align}
  \label{eq:dpd-evol-hom}
\frac{d}{d\log \mu^2}\, F(x_1,x_2, \tvec{y}; \mu)
 &= P \otimes_{x_1} F + P \otimes_{x_2} F  \,,
\end{align}
where $P$ is the well-known DGLAP splitting kernel and $\otimes_{x_i}$ the
corresponding convolution product for the variable $x_i$.  A sum over the
relevant parton species on the r.h.s.\ is understood.
Eq.~\eqref{eq:dpd-evol-hom} describes the independent evolution of the two
partons, as sketched in the right panel of Fig.~\ref{f:evol}.  A study of
the scale dependence of two-parton correlations in this framework can be
found Ref.~[\refcite{Diehl:2014vaa}] and in the presentation
[\refcite{TomasEvolution}] at this workshop.

If one integrates $F(x_1,x_2, \tvec{y}; \mu)$ over $\tvec{y}$, for
instance in the Fourier transform from $\tvec{y}$ to $\tvec{r}$, the
short-distance behavior \eqref{eq:dpd-small-y} generates a further
singularity in the form of a divergent integral $\int d\tvec{y}^2 /
\tvec{y}^2$.  An appropriate ultraviolet subtraction then leads to the
inhomogeneous evolution equation
\begin{align}
  \label{eq:dpd-evol-inh}
\frac{d}{d\log \mu^2}\, F(x_1,x_2,\tvec{r}; \mu)
 &= P \otimes_{x_1} F + P \otimes_{x_2} F 
    + \frac{f(x_1+x_2)}{x_1+x_2}\,
      K\Bigl( \frac{x_1}{x_1+x_2} \Bigr) \,,
\end{align}
which has been studied extensively in the
literature.\cite{Kirschner:1979im,%
  Shelest:1982dg,Gaunt:2009re,Ceccopieri:2014ufa} This equation has the
attractive feature that it conserves sum rules for parton number and
momentum if those are satisfied for $\tvec{r}=\tvec{0}$ at some starting
scale.\cite{Gaunt:2009re}

Which of the two evolution equations is relevant for the description of
double scattering processes can only be answered if one specifies the
factorization formula in which the distributions appear.  The homogeneous
version \eqref{eq:dpd-evol-hom} leads to a scale dependence of the
distributions that cancels the scale dependence of the hard-scattering
cross sections $\hat\sigma_1$ and $\hat\sigma_2$ in the generalization of
\eqref{eq:coll-dpd-X} to higher orders, whereas simply plugging
distributions following the inhomogeneous evolution
\eqref{eq:dpd-evol-inh} into the same formula leaves an unphysical $\mu$
dependence in the $pp$ cross section.

\begin{figure}
\centerline{
\includegraphics[width=0.36\textwidth]{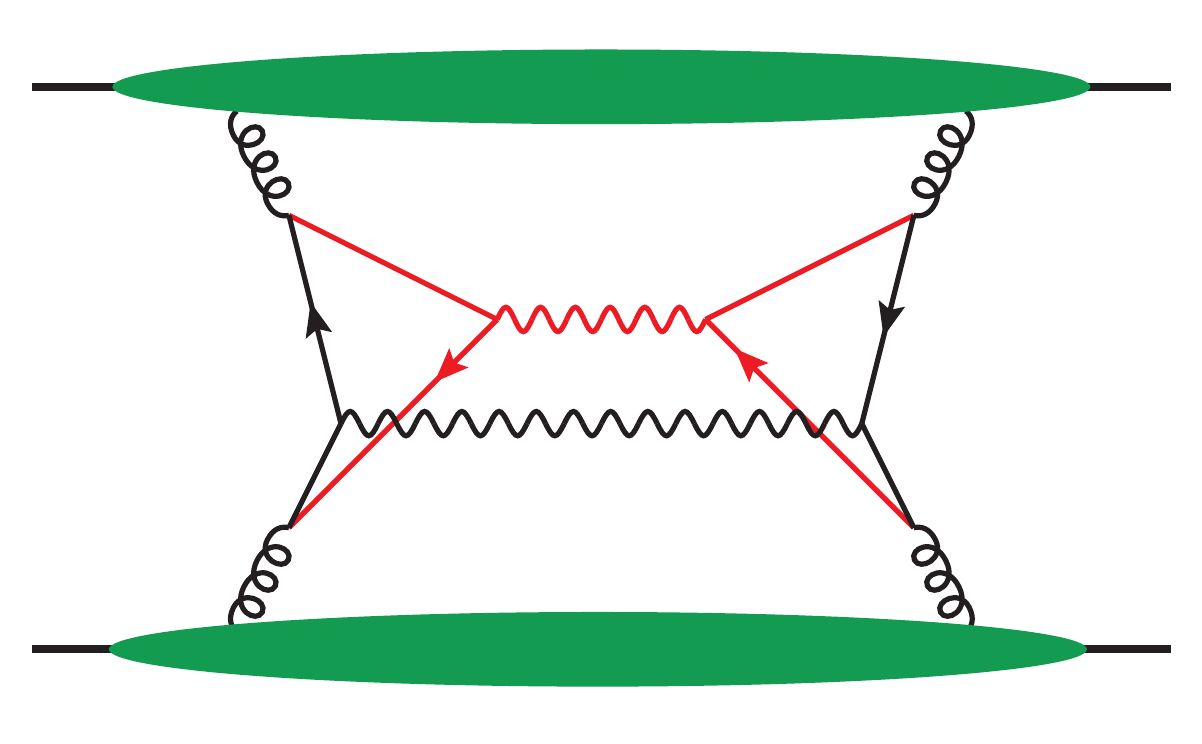}
\hspace{3em}
\includegraphics[width=0.345\textwidth,
  viewport=0 -11 330 195]{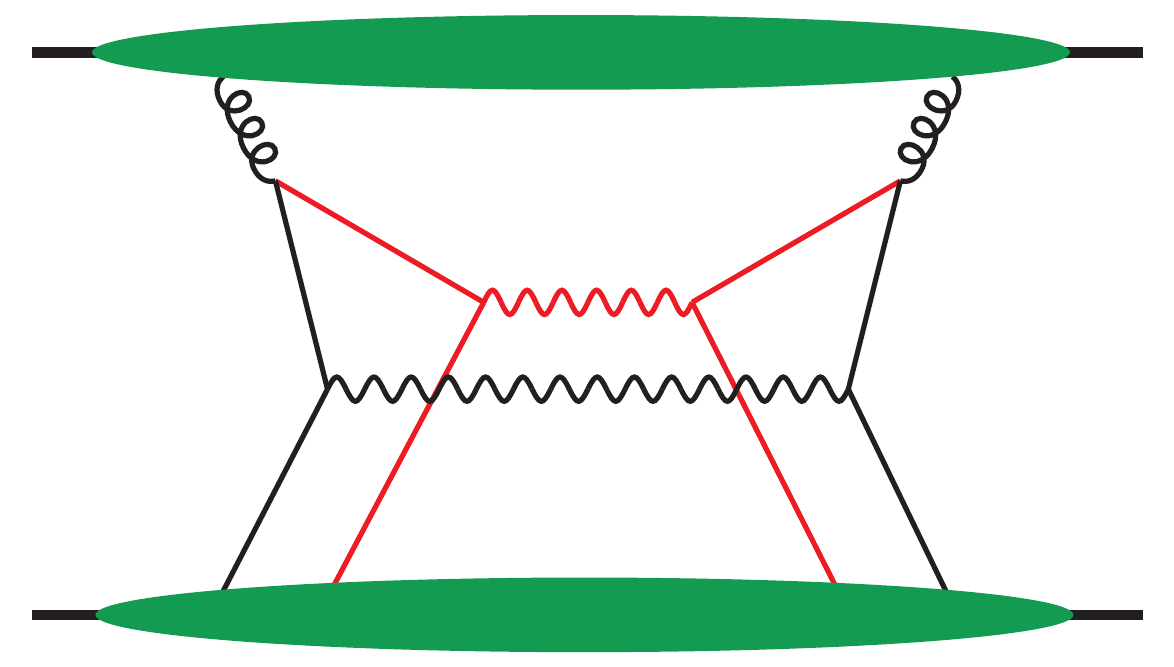}
}
\vspace*{8pt}
\caption{\label{f:split} Graphs for the production of two gauge bosons
  involving the splitting $g\to q\bar{q}$ in one or both protons.}
\end{figure}

In fact, the short-distance behavior \eqref{eq:dpd-small-y} leads to a
much more severe problem: when inserted into the factorization formula
\eqref{eq:coll-dpd-X} it gives an integral diverging like $\int
d\tvec{y}^2 /\tvec{y}^4$.  This ultraviolet divergence is connected with a
further problem, noticed already in Ref.~[\refcite{Cacciari:2009dp}] and
later exhibited in Refs.~[\refcite{Diehl:2011tt,Diehl:2011yj}].  The left
panel in Fig.~\ref{f:split} represents double hard scattering with a
perturbative splitting (cf.\ Fig.~\ref{f:evol}) in both protons, but at
the same time it can be regarded as a two-loop contribution to diboson
production by gluon-gluon fusion.  Neither in the two-loop expression for
single hard scattering nor in the double scattering formula
\eqref{eq:coll-dpd-X} is there anything that prevents double counting of
this graph in the region where the quark and antiquark coupling to a gluon
are near collinear.  A consistent description therefore requires an
appropriate modification of one or both expressions.  There is at present
no consensus on how to deal with this problem; different points of view
are discussed in Refs.~[\refcite{Diehl:2011yj,Gaunt:2011xd,%
  Blok:2011bu,Ryskin:2011kk,Ryskin:2012qx}].

A further issue in this context is the treatment of the right panel in
Fig.~\ref{f:split}, where we have a short-distance $g\to q\bar{q}$
splitting in only one of the colliding protons.  This graph and its
phenomenological relevance has been discussed in
Refs.~[\refcite{Blok:2011bu,Ryskin:2011kk,Ryskin:2012qx,Blok:2013bpa,%
  Gaunt:2012dd}].

Perhaps surprisingly, a consistent theory of double parton scattering thus
even requires us to define what exactly we mean by ``double parton
scattering'', with some amount of freedom to shift contributions between
single and double scattering terms.


\end{document}